\documentclass[runningheads]{llncs}
\usepackage{graphicx}
\usepackage{subfigure}
\usepackage{listings}
\usepackage{ xcolor} 
\begin{document}
\title{EduChain: A Blockchain-based Education Data Management System\thanks{This work was supported by Project 61902333 supported by National Natural Science Foundation of China, by the Shenzhen Institute of Artificial Intelligence and Robotics for Society (AIRS). Corresponding author: Wei Cai (caiwei@cuhk.edu.cn)}}
%
\titlerunning{Edge-assisted Federated Learning}
%
\author{Yihan Liu\inst{1} \and
Ke Li\inst{1} \and
Zihao Huang\inst{1} \and
Bowen Li\inst{1} \and
Guiyan Wang\inst{1} \and
Wei Cai\inst{1,2}}
%
\authorrunning{Y. Liu, K. Li, Z. Huang, B. Li, G. Wang, W. Cai}
%
\institute{The Chinese University of Hong Kong, Shenzhen, China \and
Shenzhen Institute of Artificial Intelligence and Robotics for Society, China \\
\email{\{yihanliu, keli, zihaohuang, bowenli,guiyanwang\}@link.cuhk.edu.cn, caiwei@cuhk.edu.cn} 
}
\maketitle              
\begin{abstract}
The predominant centralized paradigm in educational data management currently suffers from several critical issues such as vulnerability to malicious tampering, a high prevalence of diploma counterfeiting, and the onerous cost of certificate authentication. Decentralized blockchain technology, with its cutting-edge capabilities, presents a viable solution to these pervasive problems. In this paper, we illuminate the inherent limitations of existing centralized systems and introduce EduChain, a novel heterogeneous blockchain-based system for managing educational data. EduChain uniquely harnesses the strengths of both private and consortium blockchains, offering an unprecedented level of security and efficiency. In addition, we propose a robust mechanism for performing database consistency checks and error tracing. This is achieved through the implementation of a secondary consensus, employing the pt-table-checksum tool. This approach effectively addresses the prevalent issue of database mismatches. Our system demonstrates superior performance in key areas such as information verification, error traceback, and data security, thereby significantly improving the integrity and trustworthiness of educational data management. Through EduChain, we offer a powerful solution for future advancements in secure and efficient educational data management.

\end{abstract}
\section{Introduction}
Cloud services are a cornerstone of the contemporary smart campus technology landscape. Despite their prevalence, many universities still opt for centralized data management within their own data centers. This centralized approach to data handling exhibits two salient and problematic features: opacity and isolation. Opacity, in this context, refers to the lack of transparency in data access. Access is often rigidly confined to a select group of IT personnel, devoid of public oversight. This exclusivity raises concerns about potential misuse of authority, with implications such as malicious data tampering [1][13]. Isolation, on the other hand, signifies the disconnected nature of data repositories across different institutions. Data, instead of being standardized and unified, are scattered across various departments or institutions, each employing their unique standards. This lack of coherence not only impedes efficient data sharing but also inflates the cost associated with such operations. 

Additionally, this centralized data management paradigm complicates the process of academic credential verification for students. Students frequently encounter difficulties when required to authenticate their diplomas to prospective employers or higher education institutions, a process made unnecessarily complex due to the aforementioned centralization of data management. This paper explores these inherent challenges and proposes a novel, decentralized approach to educational data management.

The inherent challenges associated with the centralized approach to educational data management are threefold:

1) \textbf{Malicious Tampering}: Centralized systems can inadvertently facilitate academic dishonesty [1]. Departments such as the registrar's office often wield absolute authority over student grade information. This unchecked power can lead to potential unauthorized modifications of data, as there are no other departmental checks and balances in place.

2) \textbf{High Cost of Verification} [12]: The isolation and opacity of data across institutions give rise to substantial time and monetary costs associated with verifying students' transcripts and diplomas. The absence of a unified data management system necessitates an extensive review of numerous materials to ascertain their accuracy.

3) \textbf{Difficulty in Accountability and Error Traceback}: Conventional database systems struggle with rollback operations when errors occur, particularly in relation to aged records. This results in significant challenges in maintaining accountability and performing error tracebacks, compromising the reliability and integrity of the data.

\par
Blockchain technology presents an innovative solution to the aforementioned challenges. Gradually permeating traditional sectors such as finance, healthcare, and education, blockchain technology offers a decentralized approach to data management [1]. Essentially, a blockchain is a distributed ledger that records all transactions or digital events executed and shared among participating entities. This ledger is composed of numerous blocks, each of which is appended to the chain in chronological order. Once a block has been added to the chain, it becomes immutable. Any attempt to alter a block would incur a prohibitive cost, as each new block contains the hash value of its predecessor [8][9][14]. This inherent characteristic of blockchain technology ensures the integrity and security of the data. Transactions can be finalized without the need for a third-party guarantor, based on cryptographic principles, once consensus is achieved between participating parties. This autonomous transaction completion mechanism reduces reliance on centralized entities, enhancing efficiency and trust among participants.

\par
Traditional education blockchain systems typically employ a single-chain structure, which may prove inadequate for comprehensive information verification. In response to the escalating demands of managing educational data in complex scenarios, such as credential verification by foreign institutions, we present EduChain. EduChain is a heterogeneous blockchain system that capitalizes on the strengths of both private and consortium blockchains. The incorporation of private blockchains in EduChain ensures reliable storage for various personal information types. The consortium blockchain, on the other hand, is instrumental in recording commitments, facilitating information transfer, and verifying information authenticity. Moreover, addressing the potential issue of database mismatches across different nodes in the private blockchain, we introduce a robust mechanism for database consistency checking and error tracing. This mechanism leverages a secondary consensus via the pt-table-checksum tool, offering a comprehensive solution for maintaining data integrity in the blockchain network.

\par
This paper unfolds as follows: Section 2 embarks on a review of the relevant work, encapsulating the application of blockchain technology within educational systems, its key technological characteristics, and the landscape of existing systems. We introduce the overarching framework of our proposed system in Section 3, and Section 4 deep dives into the technical architecture, organizational logic, and functionalities. Section 5 unravels the implementation details, interspersed with illustrative code snippets for enhanced comprehension. Our system comes alive through real-time screenshots in Section 6, juxtaposed with a comparative study of competing systems. Finally, Section 7 draws together the threads of our discussion and peers into potential trajectories for future research.

\section{Related Work}
\subsection{Blockchain Technology}
The inherent properties of blockchain technology fortify the transaction environment, presenting effective solutions to traditional educational dilemmas. 1) Blockchain technology effectively mitigates malicious tampering due to its decentralized, distributed network [8][10][11]. Each modification, regarded as a transaction, is logged into the private chain and is observable by all nodes. 2) The fusion of consortium chains' transparency and private chains' confidentiality offers a means to verify students’ information while ensuring privacy. 3) Every modification is permanently logged on the blockchain, providing easy traceback, facilitating rollbacks, and enabling accountability. The technological bedrock of blockchain is the distributed consensus and the smart contract.

\subsubsection{Distributed Consensus}
Given that the blockchain database is a distributed ledger, each participant in the P2P network holds a replica of the shared database's confirmed state. Nodes maintain local data pools comprising data not yet propagated across the network. For the blockchain to be valid, the packaged data must adhere to a universally agreed standard, achieved via a consensus protocol [8]. This protocol validates blocks and standardizes data block structures. The most commonly deployed consensus protocols are proof-of-work (PoW), proof-of-stake (PoS), and delegated-proof-of-stake (DPoS)[8].

\subsubsection{Smart Contract}
Smart contracts are pivotal in eliminating the need for a third-party guarantor. These lightweight yet effective programs, stored on the blockchain, automatically execute transactions when predetermined conditions are met [9], thereby obviating the need for a third-party trust system.

\subsection{Related Blockchain-based Education System}
Blockchain technology introduces possibilities to navigate traditional educational dilemmas, prompting numerous institutions to embark on blockchain projects. Gleaning insights from these efforts significantly informs and enhances our proposed system. 

The Media Lab Learning Initiative at the Massachusetts Institute of Technology (MIT) developed a platform for creating, sharing, and verifying educational certificates [10]. While the project focuses on digitizing and verifying academic certificates, it overlooks aspects such as credit transferring. Its global reach is limited, with adoption primarily within domestic universities.

EduCTX [11] is a globally adopted platform used for issuing educational certificates and facilitating credit transfers, incorporating the European Credit Transfer and Accumulation System (ECTS). However, it lacks the capability to verify detailed academic records, rendering it susceptible to malicious tampering and diploma fabrication.

Haojian Shen and Yohuan Xiao's solution [12] addresses issues in online quizzes, such as non-transparent scoring, unfairness, and modifiable final results. While this system underscores small quizzes and exams, unlike preceding systems, it lacks the capability to verify educational certificates.
\section{System Overview}
Blockchain technology, a prominent subject in computer science, has found applications across diverse industries. Its attributes of high redundancy, immutability, traceability, transparency, and the security offered by private blockchains, are fundamental to managing educational data.

\subsection{Objective}
EduChain, when contrasted with a traditional centralized database, proposes an alternative avenue for universities to manage educational information. This system archives every modification of educational information, such as student and staff data, within a private blockchain, supported individually by each university. Every node on the private blockchain is managed by a distinct university department, and each node maintains a relational database that retains the final state of the educational information. In the event of data inconsistency among the final-state-databases across different nodes, either due to software or hardware issues, EduChain employs the online replication consistency check tool, pt-table-checksum, to execute a second consensus, identify the origin of inconsistency, and rectify it.

Beyond data storage, EduChain also facilitates swift and secure information exchange between universities and recruiters. Commitments of students’ personal information are periodically uploaded onto the consortium blockchain. In the context of our design logic, a commitment, calculated using the SHA256 algorithm, is a cryptographic primitive allowing one to commit to a chosen value or statement, keeping it concealed while retaining the ability to reveal the committed value later [2].

\subsection{Framework}
As illustrated in Fig.1, our system is structured into three primary layers: the blockchain platform layer, the front-end layer, and the user interface layer.

\begin{figure}[ht] 
\centering 
\includegraphics[scale=0.3]{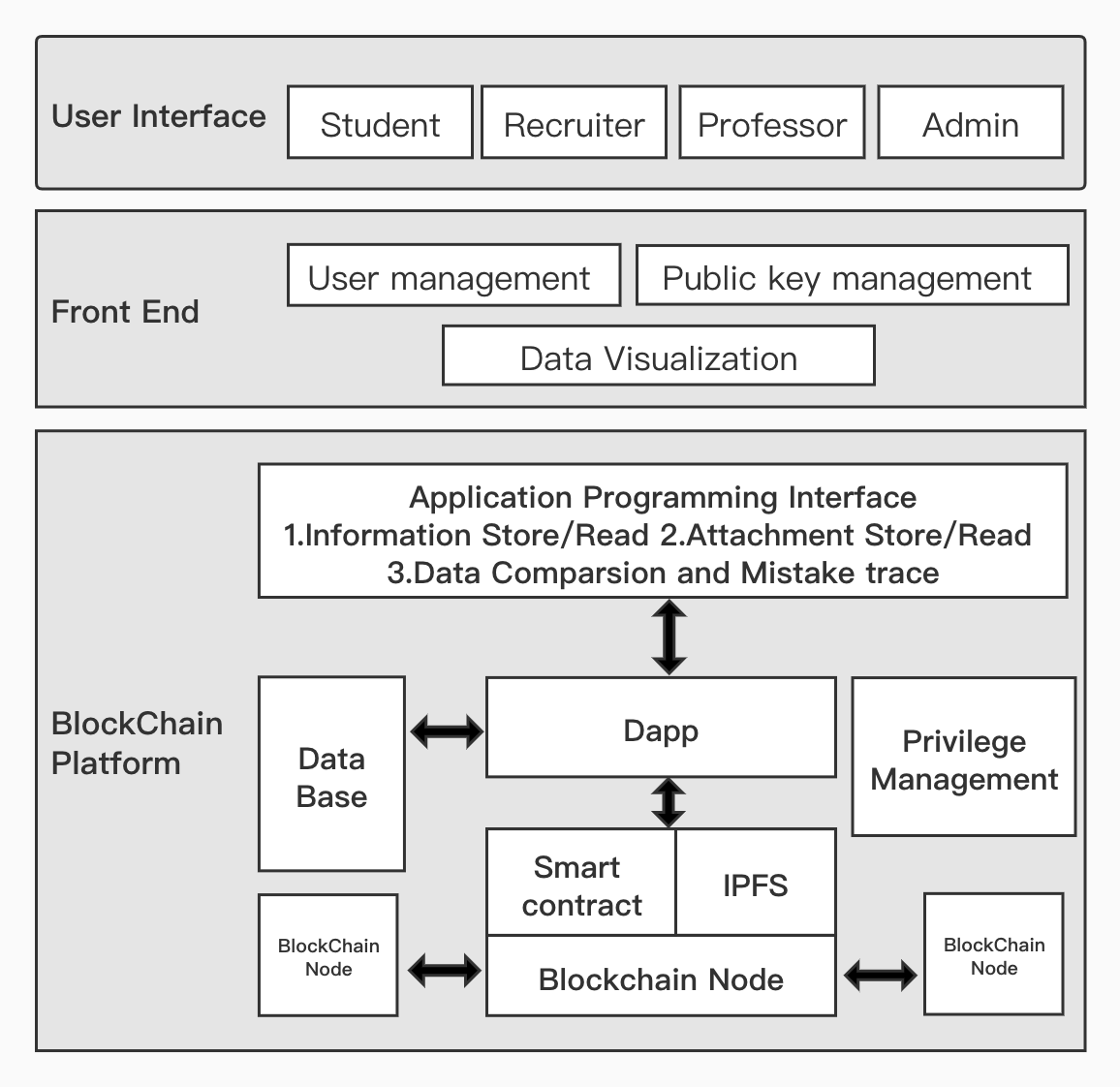} 
\caption{Framework of EduChain} 
\end{figure}

\subsubsection{Blockchain Platform}
The blockchain platform layer primarily comprises several cloud-based virtual nodes. Leveraging private and consortium blockchain technologies, all nodes are networked together. This layer offers APIs for the front-end layer to read and write data, compare data, track errors, and allows administrators to manually correct mistakes.

The blockchain platform layer is composed of one consortium blockchain and multiple private blockchains, each storing different information. The consortium blockchain is jointly maintained by the Ministry of Education and numerous universities, whereas each private blockchain is solely maintained by an individual university. All educational information is stored in the nodes of the private blockchain. Regular relational data is preserved in a relational database, while larger attachments such as photos are housed in the Inter Planetary File System (IPFS) [3], a peer-to-peer distributed file system offering secure and efficient storage for sizable attachments.

In the consortium blockchain, each university and the Ministry of Education must reach a consensus on the records in the blockchain. The consortium blockchain periodically uploads two types of information. The first includes commitments of crucial information like transcripts and diplomas. This information is accessible through a website maintained by the Ministry of Education for those who need to verify academic information, ensuring student information verification while preserving privacy. The second type of information includes requests and commitments of shared information between universities. For example, when a student wants to transfer credits from a summer session at another school, he can use the consortium blockchain. The host school sends a request to the student’s home school through the consortium blockchain using a smart contract, establishing a temporary communication channel through which the transcript is sent. This method maintains the authenticity of the transcript and protects the student's private information. The private blockchain conducts a second consensus periodically to ensure each database across different nodes is synchronized. Each node in the private blockchain also manages permission levels for different users.

\subsubsection{Front-end Layer}
The front-end layer serves as a gateway for users to access the private blockchain. It offers user management, public key management, and data visualization functionalities. The front-end layer routes requests from users across different departments to their respective nodes. After receiving responses from various nodes, this layer performs data visualization and delivers the results to the corresponding user interface.

\subsubsection{User Interface Layer}
The User Interface layer primarily deals with user interaction and page logic, interacting with the front-end layer. Staff users can use this layer to upload and modify course grades. Student users can check course information and download transcripts. All queries, modifications, and data upload requests are relayed from the front-end layer to the corresponding nodes. Any modifications and uploads require the user's signature.

\section{System Design and Functions}

\subsection{System Architecture}
Our work is principally focused on designing an education information management system, harnessing the strengths of heterogeneous blockchain technology. This system enhances the safety, reliability, and efficiency of educational information storage, sharing, and modification processes. Moreover, it provides a more secure, efficient, and transparent method of displaying student information to potential recruiters and graduate schools. Our system leverages the online replication consistency check tool, pt-table-checksum, in tandem with a secondary consensus mechanism to ensure node consensus and facilitate error tracing. 

\begin{figure}[ht] 
\centering 
\includegraphics[scale=0.11]{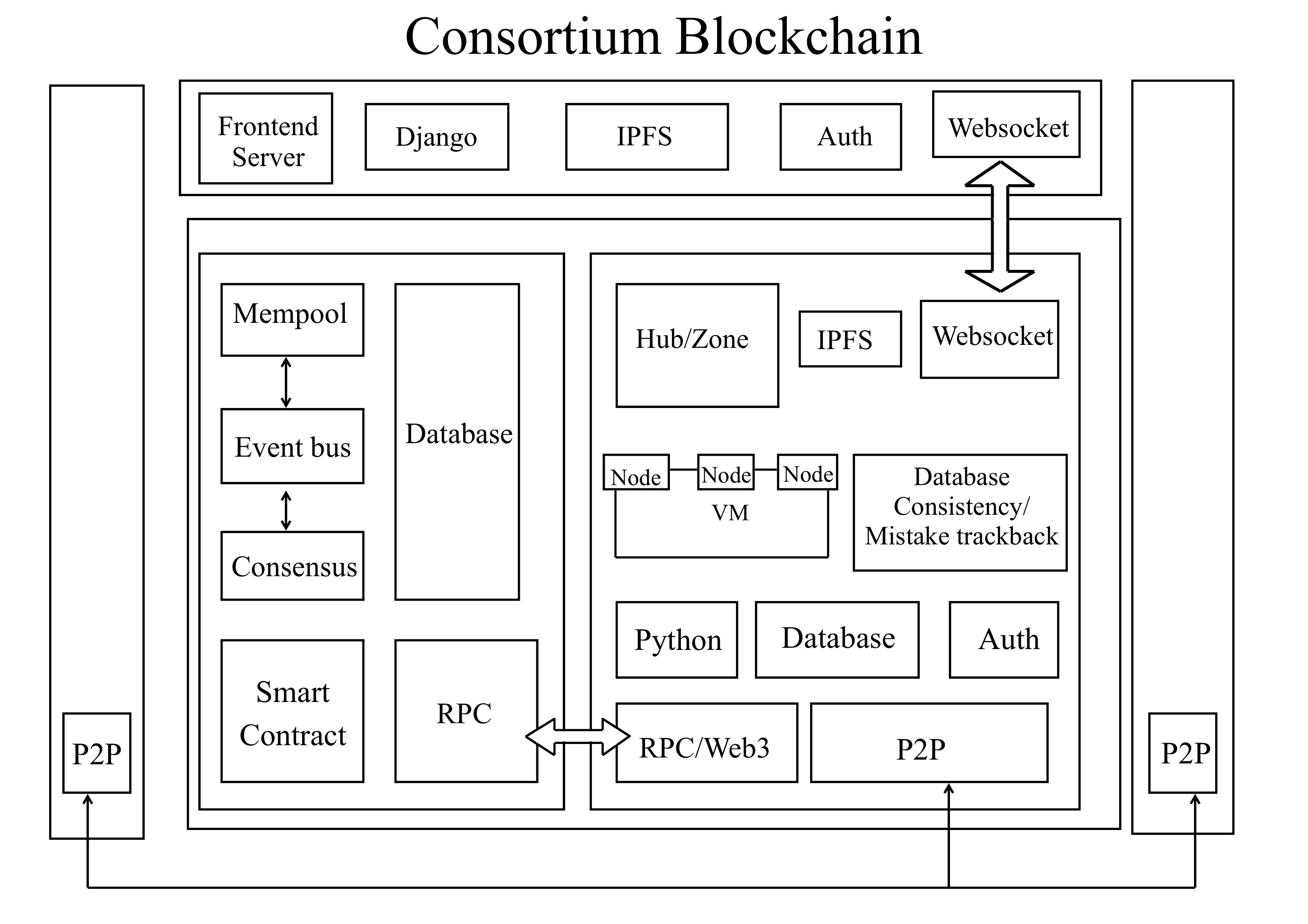} 
\caption{System design of the consortium chain} 
\end{figure}

The system architecture, as depicted in Fig. 2, employs heterogeneous blockchain technology to integrate a private blockchain (Ethereum) and a consortium blockchain (Hyperledger). The overall architecture is composed of three main components: the front-end server, private blockchain, and consortium blockchain. The private blockchain is further divided into the Distributed Application (Dapp) and the blockchain program. 

\subsection{System Components}
This section provides a brief overview of the system's primary components:

\begin{itemize}
    \item Django: A widely adopted web development framework used to construct our web server.
    \item IPFS: The InterPlanetary File System, a peer-to-peer distributed file system connecting all computing devices with a shared system of files.
    \item Auth: An identity authentication and user management module. In the web server, it verifies user identities. In a private blockchain node, it provides authority management functionality.
    \item WebSocket: A communication protocol enabling interaction between a web client and a server over the web.
    \item VM: Virtual Machines, utilizing cloud technology to construct the nodes, all of which are virtual rather than physical nodes.
    \item Database Consistency: A module that verifies the consistency of databases.
    \item Mistake Traceback: A module to identify the source of database inconsistencies across different nodes.
    \item Hub: A component that interconnects different blockchain systems.
    \item Database: Utilized to store the final state of all information, with MySQL being employed in our system.
    \item RPC: Remote Procedure Calls, a communication protocol for networked programs.
    \item web3: An Ethereum API enabling Dapps to interact with the Ethereum network.
    \item Mempool: A module temporarily storing committed transactions.
    \item Event bus: Facilitates information transfer between the mempool module and the consensus module.
    \item Consensus: A module facilitating consensus within the blockchain system. In our design, the private chain uses Proof-of-Work (PoW), while the consortium blockchain uses Kafka for consensus.
    \item Kafka: A distributed messaging system we developed for collecting and delivering high volumes of log data with low latency.
    \item PoW: Proof-of-Work, a consensus algorithm.
    \item Smart Contract: Programmed functionalities executing certain parts of legal contracts.
\end{itemize}

\subsection{System Functions}
The system can be categorized into six key functions: private chain data querying, private chain data modification, attachment uploading, consortium blockchain synchronization, verification of important student information, database consistency checking, and mistake tracing. The logic of the first three functions is shown in Fig.3.

\subsubsection{I. Private Chain Data Query}
A user's request is first sent to the front-end server. The server then relays the user's request to the corresponding nodes. The private blockchain node sends the results back to the front-end server, which performs data visualization before delivering the results to the user.

\subsubsection{II. Private Chain Data Modification}
The front-end server first receives the user's modification request and forwards it to the appropriate node, utilizing the user management module. The private node server then invokes the smart contract to perform the modification. Each node on the private blockchain uses a filter to monitor blockchain events. When a modification event is detected, the private blockchain node modifies the final state database and relays the result back to the front-end server, which then delivers the result to the user.

\subsubsection{III. Attachments Uploading}
Initially, the front-end server receives the attachment and saves it via IPFS. The server then sends the attachment reference to the corresponding nodes. The private node server invokes the smart contract to execute the upload. Upon detecting the upload event, the private blockchain node modifies the final state database and returns the result to the front-end server, which then sends it back to the user.

\begin{figure}[ht] 
\centering 
\includegraphics[scale=0.11]{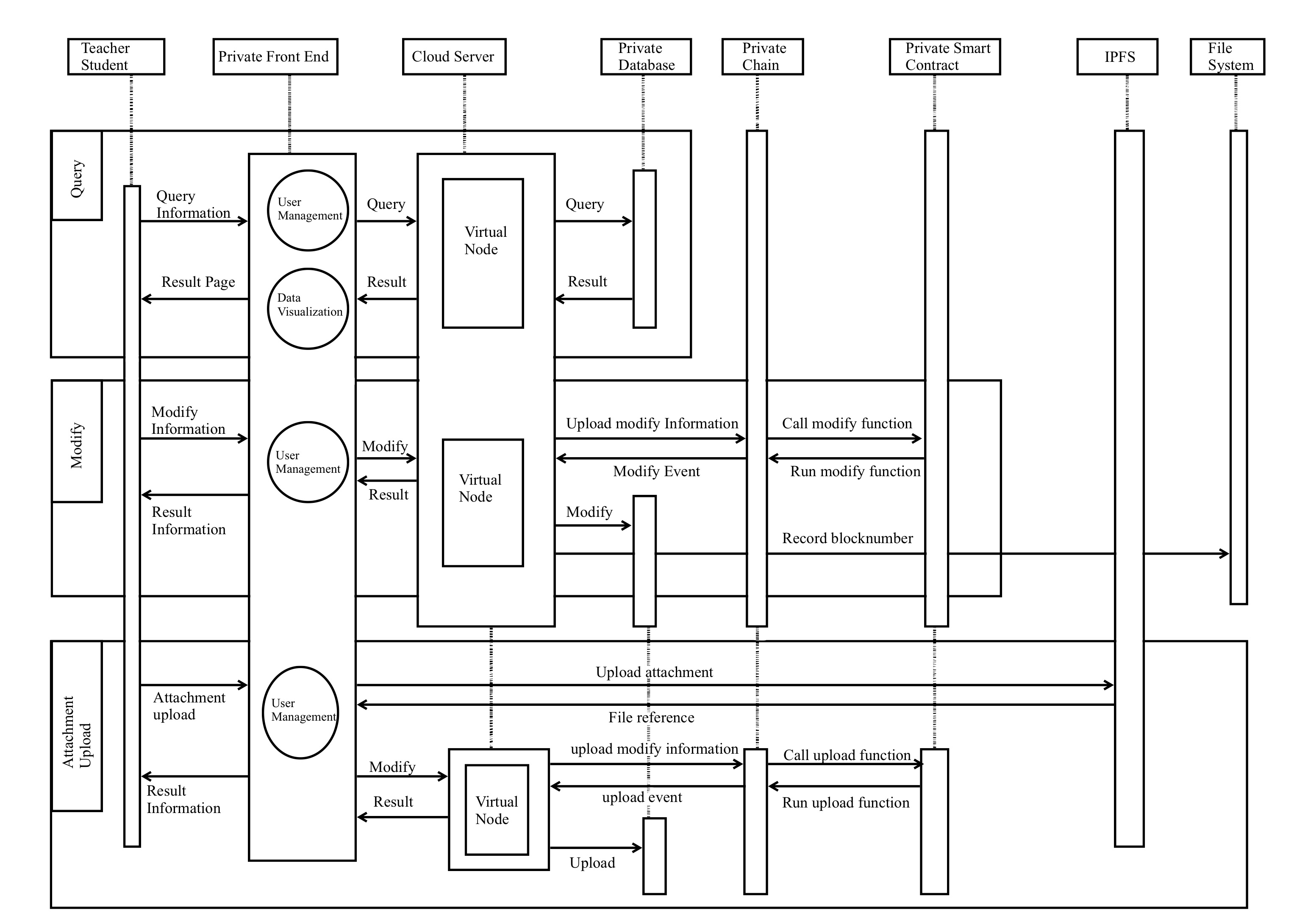} 
\caption{Logic of functions I, II, and III} 
\end{figure}

\subsubsection{IV. Consortium Blockchain Synchronization}
Each university maintains a unique node within its network. This node, not involved in typical tasks such as data retrieval, modification, or attachment uploading, acts as a hub running both private and consortium blockchains, bridging the two together.

The hub node's primary function is to transmit relevant data from the private blockchain to the consortium blockchain. It retrieves data from the final state database, transforms this data into a hash string, and invokes the smart contract in the consortium blockchain to upload the hash string.

This architecture enables seamless interaction between the private and consortium blockchains, maintaining data integrity and facilitating efficient transactions. This approach heralds future advancements in the interplay of private and consortium blockchains.

\begin{figure}[ht] 
\centering 
\includegraphics[scale=0.11]{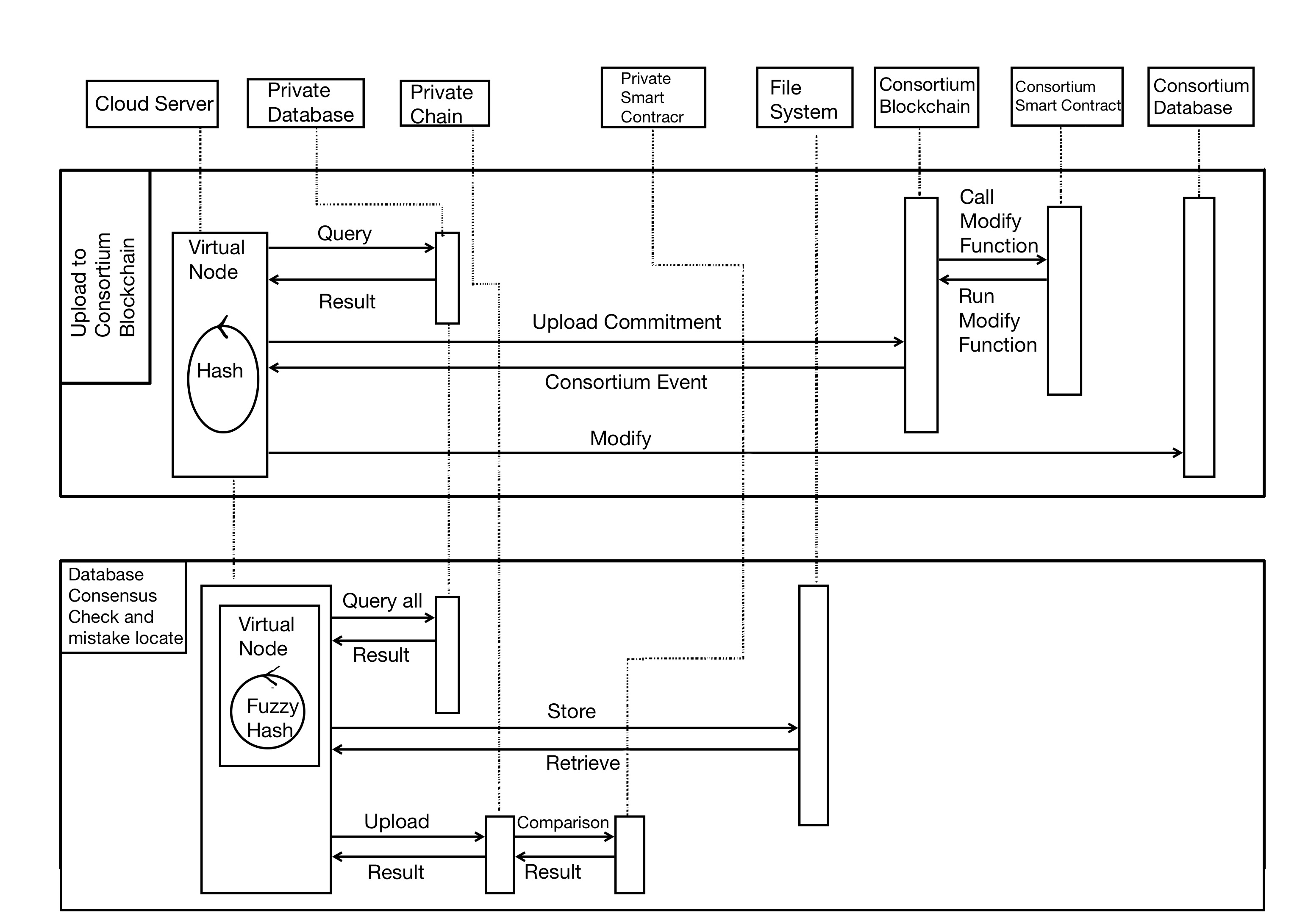} 
\caption{Logic of function IV} 
\label{fig:multi-channal GAN} 
\end{figure}

\subsubsection{V. Verification of Important Student Information}
The Ministry of Education maintains a unique node in the consortium blockchain that never connects to a private blockchain. A front-end server is built on this node. When a recruiter needs to verify certain information, they upload it to the front-end server. The node hashes the information to a string, compares the result with the hash string in the database, and sends the result back to the user.

\begin{figure}[ht] 
\centering 
\includegraphics[scale=0.2]{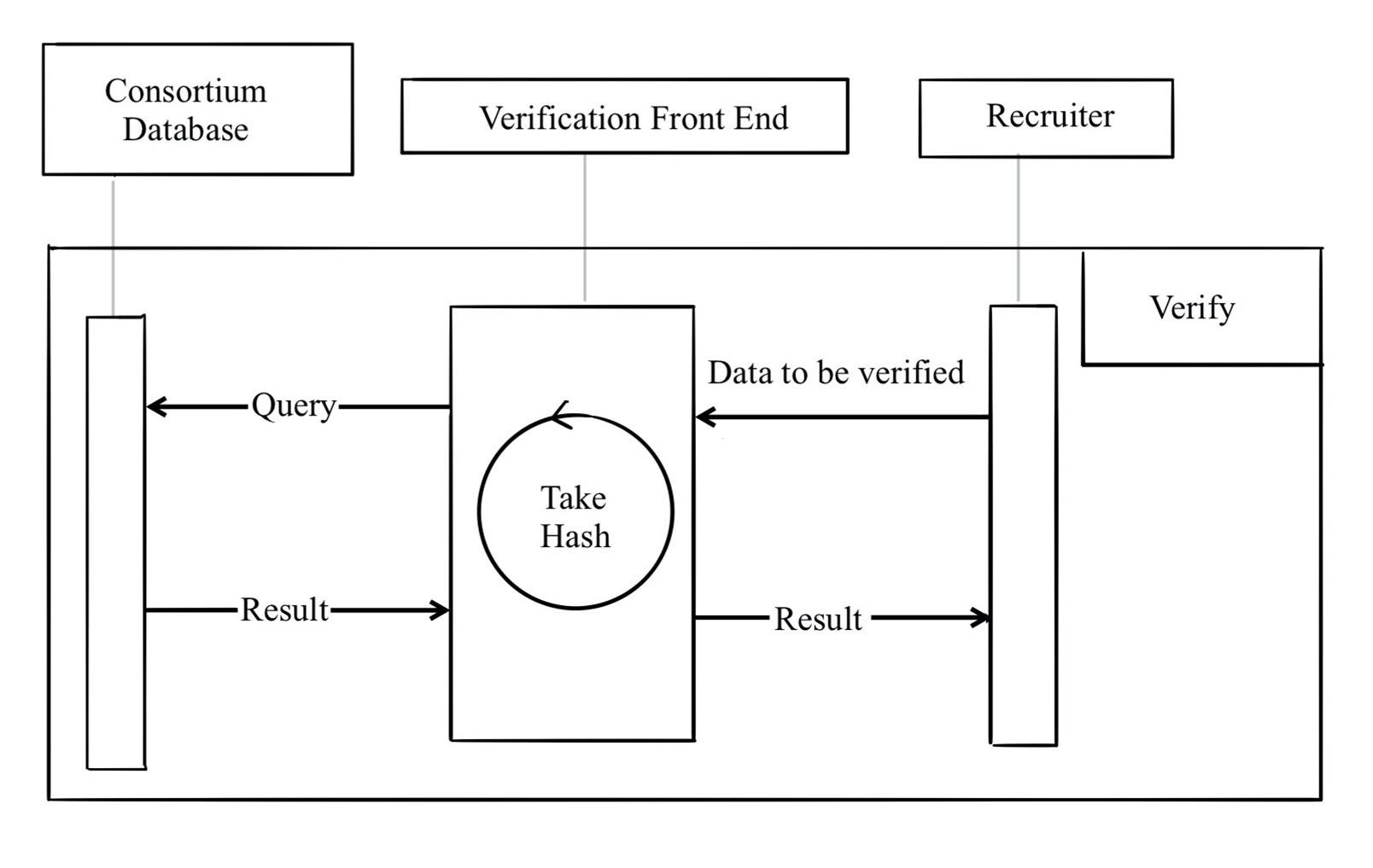} 
\caption{Logic of function V} 
\label{fig:multi-channal GAN} 
\end{figure}

\subsubsection{VI. Database Consistency Checking and Mistake Tracing}

Due to the distributed nature of the blockchain system, identical database content is dispersed across different nodes. To achieve efficient data acquisition and modification, we store the final state of each node's data in the MySQL relational database. However, this introduces a challenge: the data stored in the MySQL database diverges from the blockchain data structure, and the blockchain consensus mechanism no longer applies to these independent relational data.

To ensure data consistency, we combine the md5 verification algorithm, the pt-table-checksum MySQL master-slave database consistency detection tool, and smart contracts to implement consistency detection and error tracing of databases across different nodes. This method uses the md5 verification algorithm to compute the verification value of each database and conducts a secondary consensus by voting in a blockchain form to select the consensus database state. If a database has inconsistent data, the pt-table-checksum tool traces the error. Thanks to the characteristics of the private chain, the cost of the secondary consensus is manageable.

\section{System Implementation}

\subsection{Platform Selection}

To fulfill the functional requirements, we carefully selected an array of software packages to develop our prototypes. We used MySQL, a lightweight, free, and open-source database compatible with all mainstream platforms, for our database development. Ethereum, a decentralized open-source blockchain platform with smart contract functionality, was utilized for the private blockchain component. For the consortium chain, we opted for Hyperledger Fabric due to its robust verification mechanism, providing a secure and reliable identity verification for nodes added to the alliance chain. Given these choices, we employed Python for smart contract development on the consortium chain and Solidity for the private chain. 

To facilitate swift front-end development, we adopted the Django framework and centered our prototype development on Python. For blockchain interaction, we utilized web3.py, a Python library that offers interaction with Ethereum. For the cloud platform, we selected Amazon's VPC service, creating five EC2 virtual services to construct the network. AWS is a reputable cloud platform known for its high-quality services, making it an apt choice for prototype development.

\subsection{Prototype Deployment}

Our prototype was deployed on AWS VPC, utilizing five EC2 instances. The smart contracts were tested using the Remix online tool without running on the actual Ethereum network. Two types of smart contracts were designed: the first provided interfaces for querying and modifying related information while recording each data operation into the blockchain; the second offered interfaces for an administrator to check database consistency across all nodes and locate errors. 

For front-end deployment, we used the CodeAnyWhere online editor. Our system, following the Byzantium hard fork from the first block, was configured to meet QoS requirements and reduce latency by setting the difficulty to an extremely low level, allowing for quick new block generation. 

\lstset{
 columns=fixed,       
 numbers=left,                                        
 numberstyle=\tiny\color{gray},                       
 frame=none,                                          
 backgroundcolor=\color[RGB]{245,245,244},            
 keywordstyle=\color[RGB]{40,40,255},                 
 numberstyle=\footnotesize\color{darkgray},           
 commentstyle=\it\color[RGB]{0,96,96},                
 stringstyle=\rmfamily\slshape\color[RGB]{128,0,0},   
 showstringspaces=false,                              
 language=python,                                        
}

\begin{lstlisting}
{
  "config":{
  "byzantiumBlock":0,
  "chainID":5421,
  "homesteadBlock":0,
  "eip150Block":0,
  "eip155Block":0,
  "eip158Block":0
  },
  "coinbase":"0x0000000000000000000000000000000000000000",
  "difficulty":"0x400",
  "extraData":"0x5421",
  "gasLimit":"0xffffffffffff",
  "nonce":"0xdeadbeefdeadbeef",
  "mixhash":"0x000000000000000000000000000000000000000000
            000000000000000000000",
  "parentHash":"0x00000000000000000000000000000000000000
            00000000000000000000000000",
  "timestamp":"0x00",
  "alloc":{} 
}
\end{lstlisting}

Moreover, we set a high value for the gas limit of each block, allowing more transactions to be contained within a single block. The Ethereum network was initiated using the command shown below:

\begin{lstlisting}[language={bash}]
geth --identity "EDU" --rpc --rpcport "8545" 
--datadir data --allow-insecure-unlock --port 
"30303" --rpcapi"miner,debug,personal,db,eth,
net,web3" --networkid 5421 --maxpeers 7 console  
\end{lstlisting}

This meticulous system configuration and deployment strategy ensured that our system met all functional requirements while maintaining optimal efficiency and performance.

\section{Performance Evaluation}

\subsection{Demonstration}

\begin{figure}[htb]
\centering
\subfigure[demo1]{
\begin{minipage}[htb]{0.5\linewidth}
\centering
\includegraphics[scale=0.6]{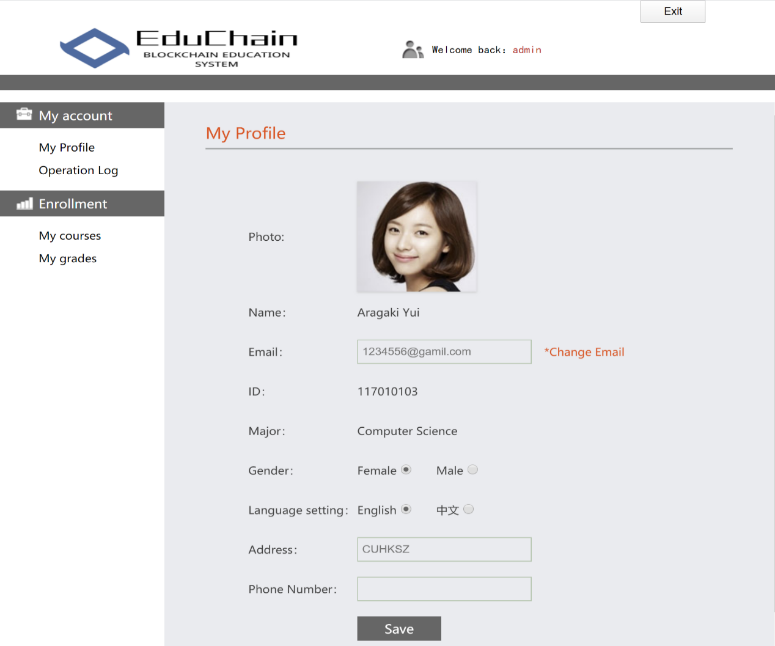}
\end{minipage}%
}%
\subfigure[demo2]{
\begin{minipage}[htb]{0.5\linewidth}
\centering
\includegraphics[scale=0.53]{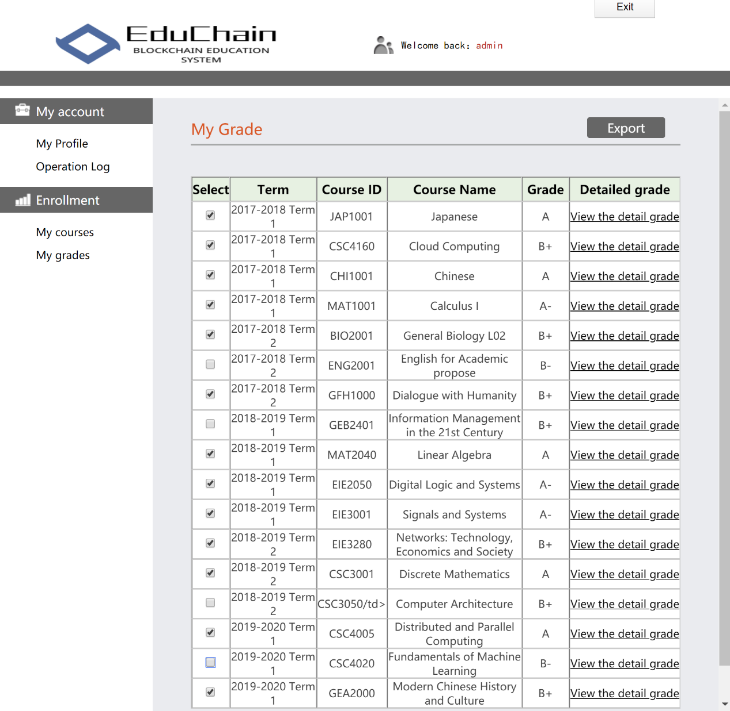}
\end{minipage}%
}%
\end{figure}

Upon successful login into the EduChain account, users can navigate to the MyProfile section via the left tool module. As illustrated in Figure 6, this page displays the user's personal information. Users can edit personal details such as telephone number, email, and address, which are stored in the corresponding node’s Inter Planetary File System and database.

Students can view their course grades on the MyGrade webpage, as depicted in Figure 7. By clicking the attached URL, students can examine detailed assessments. To export an official transcript, students must first select the relevant courses by ticking the checkboxes, then click the Export button to send a request to the central server. This process also requires the user’s password for identity verification. Upon successful identification, a download URL of the transcript is provided.

\begin{figure}[htb] 
\centering 
\includegraphics[scale=1.2]{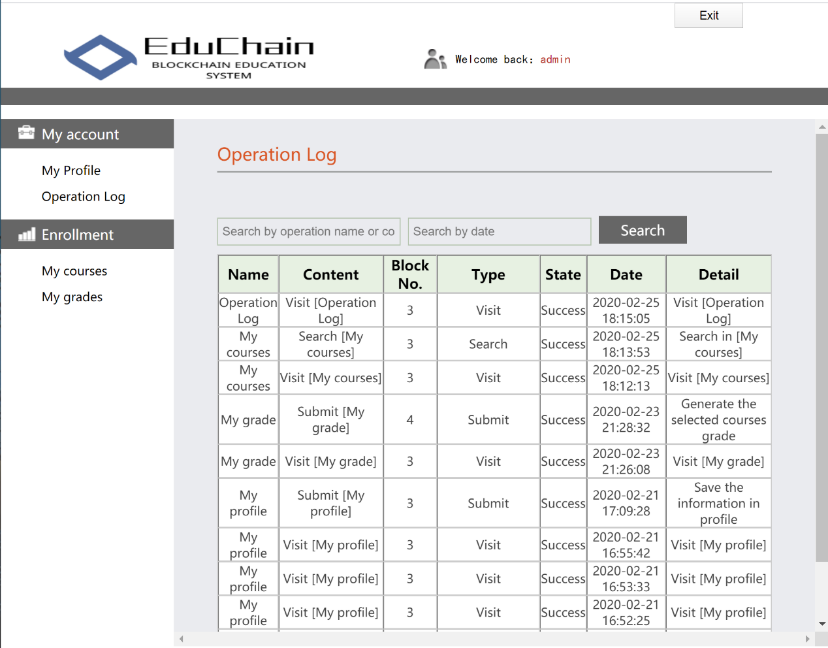} 
\caption{demo3} 
\label{fig:multi-channal GAN} 
\end{figure}

For teaching staff, EduChain allows the modification of students' final grades. Upon submission, a transaction request is sent to the server. After identification, the node server verifies the digital signature. This modification is recorded in the node's private chain, and the Dapp synchronizes the updated database with the other databases on this private chain.

Each account maintains an operation log stored in the node file system. As shown in Figure 8, we designed an operation log webpage that displays user operation details, including start time and block numbers. In conjunction with the database consistency checking and error tracing mechanism, the operation log can quickly resolve database discrepancies between nodes on the private chain.

\subsection{Comparison with Similar Systems}

\begin{figure}[ht] 
\centering 
\includegraphics[scale=0.5]{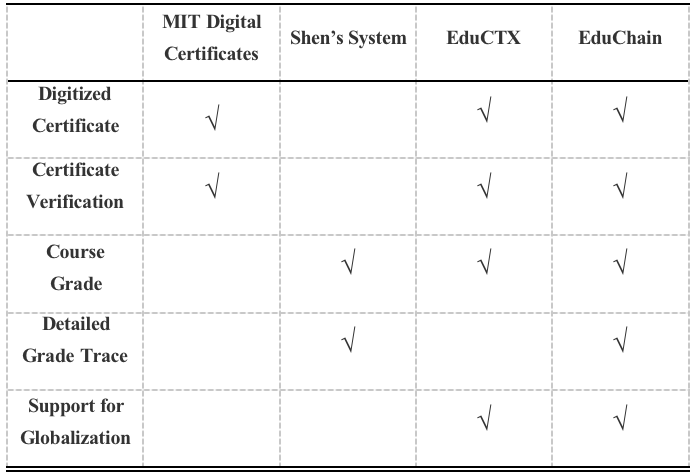} 
\caption{Comparison with MIT's, Shen's and EduCTX Systems} 
\label{fig:multi-channal GAN} 
\end{figure}

As discussed in the related work section, the MIT Digital project focuses on certificate digitization without considering global usage. EduCTX, on the other hand, is a globally accessible platform that enables certificate digitization and verification while maintaining a record of course history. However, it lacks a detailed course grade trace mechanism, making it susceptible to record fabrication before the final grade announcement. Shen's online quiz system mitigates quiz-cheating but lacks high-level functions like certificate verification. In contrast, EduChain offers digitized certificates, verification, course history information, and detailed course grades. Furthermore, the consortium blockchain underpins its global applicability.

\section{Conclusions and Future Work}
This paper presents the design and implementation of a heterogeneous blockchain system, purposed for information verification, data security, and error tracing. The system's significance lies in two key aspects: 1) By amalgamating the strengths of private and consortium blockchains, it delivers a robust and efficient method for the storage and verification of diverse information. 2) Leveraging a second consensus mechanism, we propose a rapid and effective strategy for database consistency checking and error tracing via pt-table-checksum, which proves invaluable in instances of database mismatch within the private blockchain. However, the design choice of each node maintaining its own databases gives rise to data redundancy. Future research could focus on methods to reduce data redundancy without compromising data security.



%
%
%
%

\end{document}